\documentclass[epj]{svjour}
\usepackage{graphicx}
\begin{document}
\sloppy
%
\title{Photodissociation of p-process nuclei studied by bremsstrahlung induced activation}
\author{ M. Erhard\inst{1}, A.R. Junghans\inst{1}, R. Beyer\inst{1}, E. Grosse\inst{1,2},
J. Klug\inst{1}, K. Kosev\inst{1}, C. Nair\inst{1}, N. Nankov\inst{1}, G. Rusev\inst{1},
K.D. Schilling\inst{1}, R. Schwengner\inst{1}, \and A. Wagner\inst{1} 
}                     
%
%
\institute{Forschungszentrum Rossendorf, Institut f\"ur Kern- und Hadronenphysik,
Postfach 51 01 19, 01314 Dresden, Germany
\and TU Dresden, Institut f\"ur Kern- und Teilchenphysik,
 01062 Dresden, Germany }
\date{Received: date / Revised version: date}
%
\abstract{
A research program has been started to study  experimentally  the
near-threshold photodissociation of nuclides in the
chain of cosmic heavy element production with brems\-strahlung from
the ELBE accelerator. An important prerequisite for
such studies is  good knowledge of the brems\-strahlung distribution which was
determined by measuring the photodissociation of the deuteron and by comparison
with model calculations.
First data were obtained for the astrophysically important target nucleus
$^{92}$Mo by observing the radioactive decay of the nuclides produced
by bremsstrahlung irradiation at end-point energies between 11.8~MeV and 14.0~MeV. The
results are compared to recent statistical model calculations.
} 
\authorrunning{M. Erhard, A.R. Junghans et al.}
\titlerunning{Activation of p-process nuclei...}
\maketitle
\section{Introduction}
The 35 neutron deficient stable isotopes between Se and Hg that are shielded
from the rapid neutron capture process by stable isobars,
and that are bypassed  by slow neutron captures of the s-process, are called p-process nuclei.
They are thought to be produced during supernova explosions through chains of
photodissociation reactions on heavy seed nuclei like ($\gamma$,n), ($\gamma$,p) and
 $(\gamma,\alpha)$. In proton-rich scenarios also (p,$\gamma$) reactions can occur.
The temperatures are in the region of $T=(1-3)\cdot 10^9$~K. These
temperatures need to occur on a short time scale to avoid the nuclei to be eroded by
photodissociation reactions into light nuclei in the iron region. For a current review
of the p-process see ref.~\cite{Arnould}.
In many network calculations of the p-process nucleosynthesis, Mo and Ru isotopes
are produced with lower abundances than determined experimentally.
$^{92}$Mo is the second most abundant p-nucleus with
a solar system abundance of 0.378 relative to $10^6$ Si atoms. Therefore it is
adequate to test if the photodissociation rates in the region of $^{92}$Mo that
are part of the nuclear physics input to the network calculations are correct.
We have set up an activation experiment with bremsstrahlung from the new ELBE
accelerator at Forschungszentrum Rossendorf, Dresden, to investigate the photo\-dissociation
of  $^{92}$Mo.

\label{intro}

\section{Experimental Setup}
At Forschungszentrum Rossendorf, Dresden, a new superconducting electron accelerator
named ELBE (for Electron Linear accelerator of high Brilliance and low Emittance)
has been built, which combines a high average beam current with a high duty
cycle. The accelerator delivers electron beams of energies up to 40~MeV with average
currents up to 1 mA for experiments studying photon-induced reactions. The micro-pulse
repetition rate of the accelerator can be set between 1.6 MHz and 260 MHz. In
addition, a macro pulse of 0.1 ms to 35 ms with periods of 40 ms to 1 s, respectively,
can be applied. The bremsstrahlung facility and the experimental area were designed
such that the production of neutrons and the scattering of photons from surrounding
materials is minimized~\cite{RS}. A floorplan of the brems\-strahlung facility at ELBE is shown
in Fig. 1. The primary electron beam is focussed onto a thin foil made from niobium
with various areal densities between 1.7~mg/cm$^2$ and 10~mg/cm$^2$ corresponding
to $1.6\cdot10^{-4}$ and
$1\cdot10^{-3}$ radiation lengths, respectively. After the radiator, the electron beam is
separated from the photons by a dipole magnet and dumped into a graphite cylinder
of 600 mm length and 200 mm diameter. A photo-activation site is located behind
the beam dump using available photon fluxes of up to  10$^{10}$ cm$^{-2}$ s$^{-1}$ MeV$^{-1}$.
For in-beam studies with brems\-strahlung, a photon beam is formed by a collimator made from
high-purity aluminum placed inside the concrete shielding of the accelerator hall.
Photons scattered from a target are observed by means of high-purity germanium
(HPGe) detectors. The photon flux at the target position amounts up to
10$^{8}$ cm$^{-2}$ s$^{-1}$ MeV$^{-1}$ in the bremsstrahlung cave.
All HPGe detectors are surrounded by escape-suppression
shields consisting of bismuth-germanate (BGO) scintillation detectors. In order to
determine the intensity, the spectral distribution and the degree of polarisation of the
bremsstrahlung, the photodissociation of the deuteron $^2$H($\gamma$,p)n can be used. The count
rate, the energy distribution and the azimuthal asymmetry of emitted protons is detected
by four identical silicon semiconductor detectors placed symmetrically around the
beam line at 90$^\circ$ with respect to the beam.
\begin{figure*}
\resizebox{1.0\textwidth}{!}{%
  \includegraphics{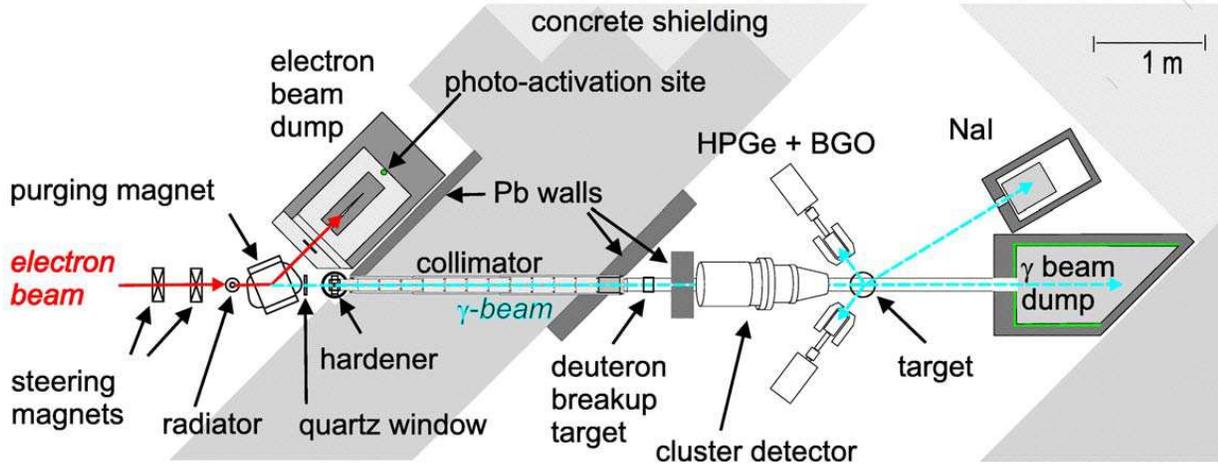}
}
\vspace*{-1cm}       
\caption{Bremsstrahlung facility and experimental area for
photon-scattering and photodissociation experiments at the ELBE accelerator.
$^{197}$Au and H${}_3$$^{11}$BO${}_3$ samples were irradiated at the target position
in the bremsstrahlung cave. $^{nat}$Mo and $^{197}$Au samples were irradiated together
at the photo-activation site.}
\label{fig:1}       
\end{figure*}

\section{Photo-activation measurements}

The number of radioactive nuclei $N_{act}(E_0)$ produced in a photo-activation measurement
is proportional to the integral of the absolute photon flux $\Phi_{\gamma}(E,E_0)$
and the photodissociation cross section $\sigma_{\gamma,x}(E)$ from the reaction threshold energy $E_{thr}$ up to
the bremsstrahlung spectrum end-point energy $E_0$.  The symbol  $x = n, p, \alpha$ stands for the
emitted particle. $N_{tar}$ is the number of the target atoms in the sample.
\begin{equation}
 N_{act}(E_0) =  N_{tar} \cdot \int_{E_{thr}}^{E_0} \sigma_{\gamma,x}(E)\cdot \Phi_{\gamma}(E,E_0)\,dE
\label{eq:yi}
\end{equation}
The number of radioactive nuclei $N_{act}(E_0)$ is determined experimentally after irradiation
with bremsstrahlung in a low-level gamma-counting setup using a 100$\%$ HPGe detector:
\begin{equation}
 N_{act}(E_0) = Y(E_{\gamma}) \cdot \kappa_{corr} / \big( \varepsilon(E_{\gamma}) \cdot p(E_{\gamma}) \big)
\label{eq:yint}
\end{equation}
$ Y(E_{\gamma}),\varepsilon(E_{\gamma}), p(E_\gamma)$ are the dead-time and pile-up corrected
full-energy peak counts of the observed transition, the absolute efficiency of the detector at the energy
$E_{\gamma}$ and the emission probability of
the photon with energy $E_\gamma$.
 The factor $\kappa_{corr}$ contains
the relation of the detected decays in the measurement time $t_{meas}$
to the number of radioactive nuclei present.
Decay losses in the time $t_{loss}$ in between the bremsstrahlung irradiation
and the begin of the measurement as well as decay during the irradiation time $t_{irr}$ are
taken into account. The symbol $\tau$ denotes the mean life time of the radioactive nucleus produced
during the photo-activation.
\begin{eqnarray}
 \kappa_{corr} = \frac{\exp(t_{loss}/\tau)}{1 - \exp(-t_{meas}/\tau)} \cdot 
  \frac{t_{irr}/\tau}{1 - \exp(-t_{irr}/\tau)}
\end{eqnarray}
The constancy of the electron beam current and thus of the photoactivation
rate was checked by monitoring both the electron current in the injector and in the beam dump.
During a typical irradiation time of 8-16 hours there were no electron beam outages.

The set up can be used for photo-activation measurements in the following way:
The sample (Mo) is activated in the high photon flux behind the beam dump together with a Au sample
to measure an activation standard reaction, e.g.  $^{197}$Au($\gamma$,n)
(photo-activation site in Fig.~\ref{fig:1}).
During the same experiment another Au sample is irradiated at the target
position inside the bremsstrahlung cave. There, the absolute
photon flux can be determined from the  ($\gamma,\gamma'$) yield of a sample containing $^{11}$B which is
irradiated in the same place. In $^{11}B$ the ground state transition width of 4 levels is known
with sufficient accuracy~\cite{TUNL}.
This measurement is done during the entire activation period
with HPGe detectors. The cross section of $^{197}$Au($\gamma$,n)
is then renormalized to give the measured activation yield with the
absolute photon flux determined experimentally.
With the renormalized  $^{197}$Au($\gamma$,n) cross section and a simulated thick target
bremsstrahlung spectrum, the absolute photon flux at the photo-activation site behind the beam
dump can be determined. From the absolute photon flux and the measured activation yield
the cross section normalization for photodissociation of p-process nuclei like
$^{92}$Mo can be determined.

\subsection{Photon spectrum and end-point energy determination}
\begin{figure}

\resizebox{0.43\textwidth}{!}{%
  \includegraphics[angle=0] {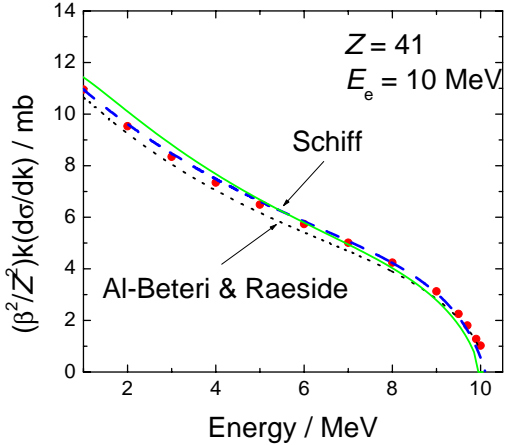}
}

\caption{Theoretical bremstrahlung cross sections in comparison
with the evaluation of Seltzer and Berger (circles)\cite{BS} for a Nb radiator
and electron kinetic energy of 10~MeV.  The full line is from ref.~\cite{Roche} the
dashed line is from ref.~\cite{BH}. They were calculated using a program from
E.~Haug~\cite{Haug} that also includes a screening correction according to ref.~\cite{Salvat}.
The dotted line is from the parametrization~\cite{Albeteri}.}
\label{fig:2}
\end{figure}

The bremsstrahlung spectrum at the target position can be described well based on the brems\-strahlung
cross section of a thin target.  Fig.~\ref{fig:2} shows theoretical brems\-strahlung cross sections
compared to the evaluation of Seltzer and Berger \cite{BS} for a niobium radiator.
The calculations were made using formulae given by Schiff~\cite{Schiff}, Heitler~\cite{BH},
and Roche~\cite{Roche}; the last one being corrected and programmed by Haug~\cite{Haug}, who
also included an updated screening correction due to Salvat et al.~\cite{Salvat}.
Al Beteri et al. performed a theory motivated parametrization of experimental data as known
in 1988~\cite{Albeteri}.

All curves shown give results that agree to within 5 percent of each other at the low
energy side of the spectrum, when atomic screening is taken into account.
At the high energy side at about $\simeq$1~MeV below the end point the descriptions
differ to  typically 20 percent. This will influence the activation yield
around the reaction threshold, where the yield integral
depends strongly on the overlap with the high-energy tail of the photon spectrum.
For $^{197}$Au($\gamma$,n), e.g., (cross section calculated with ref.~\cite{Talys})
 the yield integral calculated with the Al-Beteri cross section
\cite{Albeteri} is about 25 percent higher at 100 keV above $E_{thr}$, compared to
using the formula given by Schiff~\cite{Schiff}. At  $E_{thr}$~+~1.5~MeV  the uncertainty due to the
different photon spectra is below 5 percent.

The electron beam energy and the electron beam energy width determine the
high-energy part of the brems\-strahlung spectrum. They need to be known with high precision.
An uncertainty of the electron beam energy of
$\pm$100 keV at energies around 10~MeV has significant influence on the yield integral.
For $^{197}$Au($\gamma$,n) this error would change the yield integral by 20(10) percent
at end-point  energy $E_0$ = $E_{thr}$ + 1.0(2.0) MeV. The data shown in this work were measured
at $E_0  >$ 11.8~MeV which is several MeV above the respective $^{197}$Au($\gamma$,n) and
$^{92}$Mo($\gamma$,p) reaction thresholds.
The electron beam energy was determined through the ion optical setting of the
accelerator and the beam line and also on-line by measuring the deuteron breakup.

To a lesser extent the absolute photon flux in the experiment also depends on the
electron beam energy width, which needs to be taken into account in the photon flux determination for
reaction yields measured close to the reaction threshold. At ELBE, the energy width of the beam has been measured ion optically to be 60 keV (FWHM) during the measurements discussed here.

\begin{figure}

\resizebox{0.43\textwidth}{!}{%
  \includegraphics[angle=0] {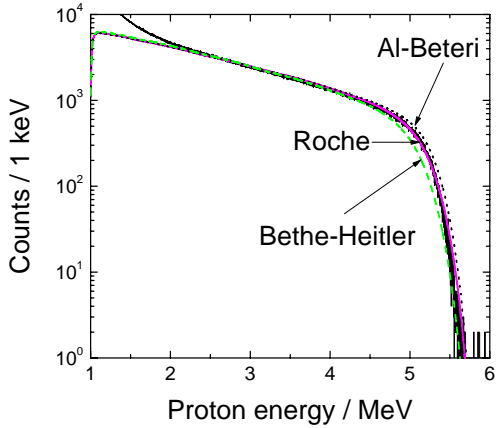}
}

\caption{Proton kinetic energy spectrum from the photodissociation of a
deuterated PE target (histogram) compared with simulated spectra based on the
Al-Beteri (dotted), Roche (full), Bethe-Heitler (dashed); references, see Fig.~\ref{fig:2}}.
\label{fig:4}
\end{figure}

We have determined experimentally the
end-point  energy and also the spectral distribution of the bremsstrahlung
by measuring proton spectra from deuteron breakup, shown in Fig.~\ref{fig:4}. For details,
see ref.~\cite{RS}.
The target is a  thin deuterated polyethylene  foil (areal density 4 mg/cm$^2$).
Caused by the extended target size, the Si-detectors register
protons emitted at different angles causing a kinematic spreading of the proton energy
distribution of 150 - 200 keV.  This spread dominates the resolution of the simulated
proton spectrum shown in Fig.~\ref{fig:4}.   The simulated proton spectra are based on
different bremsstrahlung cross section formulae. 
They include the energy resolution of the ELBE beam and the passage of 
the bremstrahlung through the Al-hardener in front of the collimator. 
Deviations below 2~MeV are due to photon and electron background in the
silicon detectors. The uncertainty in the end-point  energy determination is $\pm$ 100 keV.
To improve the accuracy of the measurements
we have moved the detectors further away from the target foil
to reduce the effect of the reaction kinematics.
Fig.~\ref{fig:4} shows that the spectrum can be described well with the Roche or Al-Beteri
cross sections, whereas the Bethe-Heitler formula is lower, as Coulomb correction
terms are not included there.

\begin{figure}

\resizebox{0.43\textwidth}{!}{%
  \includegraphics[angle=0] {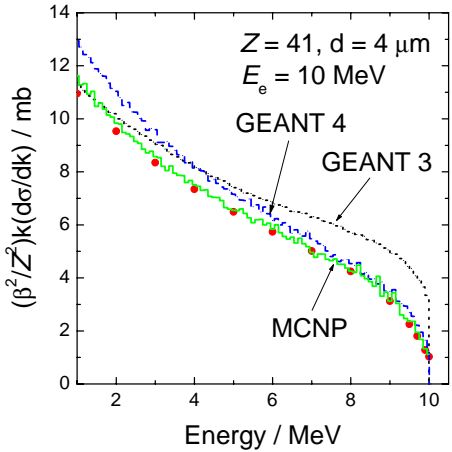}
}
\caption{Monte Carlo simulations~\cite{MC} of bremsstrahlung spectra in comparison
with the evaluation of Seltzer and Berger~\cite{BS} (circles) for a Nb radiator
and electron end-point  energy of 10~MeV. The dotted histogram is calculated with GEANT3,
the dashed histogram with GEANT4. The full histogram is calculated with MCNP4C2.}
\label{fig:3}
\end{figure}

We have also compared different Monte-Carlo Simulations to the tables of Seltzer and Berger, by
making simple simulations of the absolute photon spectrum created in  a thin Nb radiator.
Fig.~\ref{fig:3} shows that both versions of GEANT~\cite{MC} 
show appreciable differences from the Seltzer and Berger evaluation, 
while the MCNP 4C2 (Monte Carlo N-Particle Transport Code)~\cite{MC}
simulation corresponds very closely to the data.
MCNP uses the Seltzer and Berger evaluation~\cite{BS}. For GEANT3 the CERN Program library
long write up states that Seltzer and Berger is used as well.
GEANT4 takes bremsstrahlung cross section input from the Evaluated Electron Data Library~\cite{EEDL}. 
The simulations from GEANT4 and MCNP agree only at energies above 7~MeV when
using 10~MeV electrons.
From our study of the deuteron breakup we strongly favour the Roche code \cite{Roche,Haug,Salvat}
and the concurrent tables of Seltzer and Berger~\cite{BS}, resp. the MCNP4C2 code.

\begin{figure}

\resizebox{0.45\textwidth}{!}{%
  \includegraphics[angle=270] {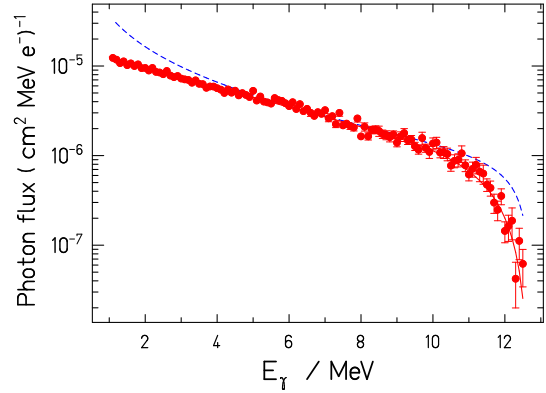}
}

\caption{Bremsstrahlung spectrum at the main photo-activation site behind the
graphite beam dump as calculated with MCNP 4C2 (circles) for an end-point energy of
12.6~MeV. The dashed line denotes a
theoretical thin target bremsstrahlung spectrum calculated according to Schiff \cite{Schiff}.
It is normalized to the MCNP simulation at 6~MeV. The full line is a parametrization
of the MCNP simulation. }
\label{fig:5}
\end{figure}

With MCNP4C2 we have calculated the absolute photon flux at the photo-activation site
behind the beam dump, see Fig.~\ref{fig:5}. From the comparison with a thin target Schiff
spectrum one can see how the shape of  a thick target spectrum is changed due to
creation of photon-electron cascades and multiple scattering.
The thick target spectral shape is required to determine the absolute photon flux
for the Mo samples that were irradiated at the photo-activation site with the bremsstrahlung
produced in a thick graphite block. 
Based on a parametrization of the MCNP results the thick-target photon flux was calculated that was
used in the analysis of our Mo photo-activation data.

\subsection{$^{197}$Au($\gamma$,n) as activation standard}

$^{197}$Au samples of approx. 200 mg have been irradiated at the target position together
with an H${}_3$$^{11}$BO${}_3$ sample enriched in  $^{11}$B to 99.27 percent and mass 2.93 g.
Activation measurements were performed at end-point  energies of 11.8~MeV up to 14.0~MeV
The absolute photon flux was determined using known  transition strengths
in $^{11}$B at $E_{\gamma}(\Theta_{lab} = 90^{\circ})$ = 4444, 5019, 7283 and 8916 keV, respectively. Detectors were positioned at
$\Theta_{lab} = $ 90$^{\circ}$ and $\Theta_{lab} = $ 127$^{\circ}$.
Angular correlation effects are important for the $^{11}$B photon scattering yields as
observed especially at 90$^{\circ}$. Feeding corrections are
estimated to be small for the high-energetic transitions (no feeding for the highest
transition) used here, but will be included in the final data analysis.

The number of $^{196}$Au nuclei produced during the activation was determined from
decay measurements in a low level counting set up with a 100\% HPGe detector. The total efficiency
was measured with the help of several calibration sources from Amersham and
PTB~\cite{PTB} to 3 percent uncertainty in the energy range from 150 - 2000 keV. A Cd absorber was used
to minimize coincidence summing effects with x-rays emitted from the Au samples and some of the
calibration sources.
The efficiency was checked by GEANT3 simulations that were adjusted to the experimental data to
give the efficiency as a function of photon energy.
Coincidence summing corrections for the 333 keV and 356 keV decay lines of $^{196}$Au
were taken into account. The weaker transition at 426 keV into Hg
that does not have coincidence summing was included in the analysis.
The number of nuclei produced during the activation at the target position was normalized
to the number of $^{197}$Au target nuclei and to the photon flux at $E_{\gamma}$ = 8916 keV.

 \begin{figure}

\resizebox{0.5\textwidth}{!}{%
  \includegraphics[ angle=0] {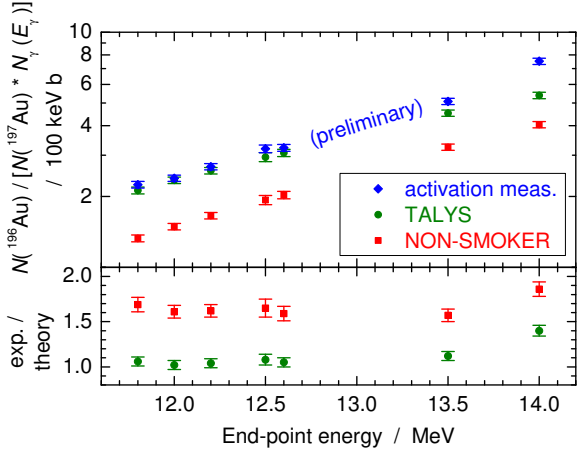}
}
\caption{Preliminary activation yield of $^{197}$Au($\gamma$,n)
      measured at the target position, see Fig.~\ref{fig:1}.
      The experimental yield is normalized to the number of $^{197}$Au atoms and to the
      absolute photon flux at the energy $E_\gamma =$ 8916 keV.
      The data are compared to yield integrals computed with the cross sections from
TALYS and NON-SMOKER using the absolute photon flux determined from known transitions in
a sample containing $^{11}$B. }
\label{fig:6}
\end{figure}

The data for $^{197}$Au($\gamma$,n) are shown in Fig.~\ref{fig:6} in comparison with results that were calculated using
the absolute photon flux determined experimentally and the theoretical
 $^{197}$Au($\gamma$,n) cross section from the TALYS program~\cite{Talys} and from NON-SMOKER
\cite{Rauscher}. In the range below 13~MeV the theoretical result from TALYS  is about 10  percent lower
than the experimental values. The combined effect of the  systematic uncertainties in the
absolute photon flux related to the end-point  energy, spectral shape of bremsstrahlung and electron-beam energy resolution,
as discussed above have not been finally determined yet. About 20 percent uncertainty does seem
to be realistic, however. From the data shown in Fig.~\ref{fig:6} we conclude that the
results from the NON-SMOKER code are considerably lower than observed experimentally.
The yield integrals calculated with NON-SMOKER to end-point  energies from 11.8~MeV to 14~MeV
are only 60 to 80 percent of the yield integrals calculated with TALYS.

\begin{figure}
 \resizebox{0.43\textwidth}{!}{%
  \includegraphics[angle=270] {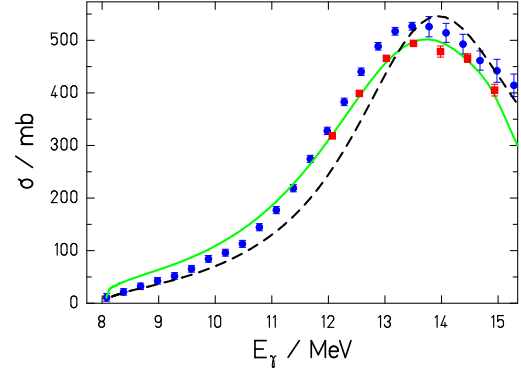}
}

\caption{Measured  $^{197}$Au($\gamma$,n) from positron annihilation in flight compared
with two model calculations. The squares denote the data from Berman et al., the circles
data from Veyssiere et al.. The dashed line is the prediction from Rauscher and Thielemann
\cite{Rauscher},
whereas the full line is calculated using the TALYS code from Koning et al.~\cite{Talys}.
 For references, see \cite{sumcit}. }

\label{fig:8}
\end{figure}

To investigate the discrepancy of the models, we compare in Fig.~\ref{fig:8}  $^{197}$Au($\gamma$,n) cross section data
obtained with quasi-monoenergetic photons from positron annihilation in flight in comparison with the model calculations.
Up to $E_{\gamma}$ = 13~MeV the predictions from the NON-SMOKER code are systematically lower.
TALYS   closely matches the experimental data  around the peak region of the GDR.
The tails of the GDR are not described well by either model, therefore it is not straightforward
to use $^{197}$Au($\gamma$,n) as an activation standard close to the reaction threshold around
8~MeV, as was also realized previously~\cite{Vogt}.

\subsection{Photo-activation measurements of Mo-Isotopes}

\begin{figure}
 \resizebox{0.45\textwidth}{!}{%
  \includegraphics[angle=270] {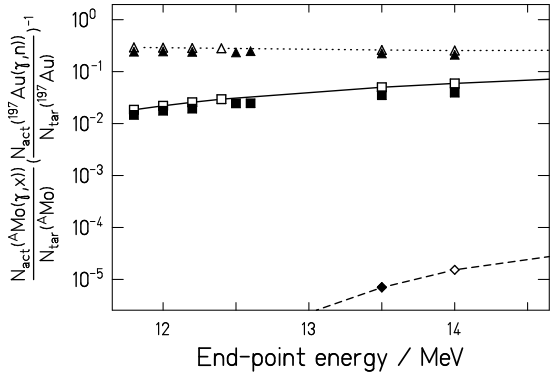}
}

\caption{ Measured activation yields for different Mo-Isotopes at the photo-activation site as
 a function of the bremsstrahlung end-point  energy.
The data are normalized to the activation yield from $^{197}$Au($\gamma$,n) irradiated simultaneously.
The full symbols denote the experimental yields of $^{100}$Mo($\gamma$,n) (triangles),
$^{92}$Mo($\gamma$,p) + ($\gamma$,n)  (squares), and  $^{92}$Mo($\gamma,\alpha$) (diamond).
The effect of different target masses is  taken into account. The open symbols connected
with lines to guide the eye represent yield integrals calculated with the photodissociation
cross sections from Rauscher and Thielemann \cite{Rauscher}. The yield integrals of the
reactions involving the Mo isotopes are divided by the $^{197}$Au($\gamma$,n) yield integral. }
\label{fig:7}
\end{figure}

Natural samples of molybdenum (mass 2 - 4 g, disc diameter 20 mm) have been irradiated
together with the Au samples as discussed above. We also did measurements with enriched
Mo samples to study the dipole strength around the particle threshold, see ref.~\cite{EGrosse}.
With an enriched $^{92}$Mo sample we observed the  $^{92}$Mo($\gamma,\alpha$)
reaction at the rather low end-point energy of 13.5~MeV.
As the absolute normalization of the photon flux at the photo-activation site is still in progress,
Fig.~\ref{fig:7} shows the measured reaction yields relative to the experimental Au reaction yield
as calculated in eq.~(\ref{eq:yint}). The data are normalized to the different number of target atoms in
the samples. The experimental data points are compared with the yield integrals calculated
with the simulated thick-target bremsstrahlung spectrum shown in Fig.~\ref{fig:5} and the
NON-SMOKER photodissociation cross sections. The yield integrals are calculated relative to the
$^{197}$Au($\gamma$,n) yield integral. The $^{92}$Mo($\gamma,\alpha$) data point
is taken relative to the $^{92}$Mo($\gamma$,p) yield integral measured with the same target.

The data agree on a scale relative to $^{197}$Au($\gamma$,n) to typically 20 - 30 percent with
the simulation. One can see from these measurements,
 that the  ($\gamma$,p) reaction cross section for the neutron-deficient isotope $^{92}$Mo
has about the same size as the ($\gamma$,n) and extends to lower energies.
 The $^{92}$Mo($\gamma$,p) reaction cross section is dominant at energies below
12.6~MeV, as the   $^{92}$Mo($\gamma$,n)   channel is not open yet.  The $^{91}$Nb nuclei produced
were identified by the 1205~keV transition following the $\beta$ decay into $^{91}$Zr.
 The population of the long lived
($t_{1/2} = 680$ yr) ground state in $^{91}$Nb that cannot be easily detected in an activation measurement
has to be taken into account, e.g. by statistical model calculations. With the TALYS code we have
calculated that this effect is around 5 - 10 percent in the energy range measured.
The contribution from  $^{92}$Mo($\gamma$,n) above 12.6~MeV produces shorter lived activity
that can be identified with a pneumatic delivery system that is under developement.
The $^{100}$Mo($\gamma$,n) results will be used for comparison with Coulomb dissociation
experiments done at GSI, Darmstadt~\cite{KS}.
The similarity of the relative data shown in Fig.~\ref{fig:7} as compared to the 
NON-SMOKER data~\cite{Rauscher}
suggest already that the predicted underproduction of Mo/Ru isotopes might not be due to wrong photodissociation
rates. Absolute data currently under analysis will allow to draw a firm conclusion.

\section{Conclusion}
First photodissociation measurements of the p-process nucleus $^{92}$Mo have been performed
at the new bremsstrahlung experiment at FZ Rossendorf, Dresden. The activation technique has
been used to identify the different reaction products. The photodissociation reactions
($\gamma$,n), ($\gamma$,p), and ($\gamma,\alpha$)  have been observed. The bremsstrahlung spectrum has been studied
using the photodissociation of the deuteron. This allowed to measure the end-point  energy of the
electron beam and to compare different formulae for thin-target bremsstrahlung.
The absolute photon flux has been measured online by known transitions in $^{11}$B and by
using the reaction $^{197}$Au($\gamma$,n) as an activation standard.
Preliminary results indicate that the absolute cross section of  $^{197}$Au($\gamma$,n)
as calculated by the NON-SMOKER code is lower than the experimental data. This was also
found in the energy region below 10~MeV in ref.~\cite{Vogt}.
The reaction yields for $^{92}$Mo($\gamma$,p) + ($\gamma$,n)  relative to  $^{197}$Au($\gamma$,n)
agree to within 20 - 30 percent with the model calculations~\cite{Rauscher}.
These reactions contribute to the possible destruction of the p-process nucleus $^{92}$Mo.

\section*{Acknowledgments}
We thank  P. Michel and the ELBE-crew for providing stable, high-intensity electron beams
for the activation measurements and A. Hartmann and W. Schulze for continuous valuable
technical support. We gratefully acknowledge theory discussions and help by H.W. Barz and
E. Haug.

%
%

\end{document}